# ON SURFACE PLASMON SPECTRUM IN NOBLE METAL NANOPARTICLES – RODS AND SPHEROIDS

*A.Melikyan and H.Minassian*

*State Engineering University of Armenia, Yerevan, 375009 Armenia*

The surface plasmon oscillations spectra in noble metal nanoparticles - rods and spheroids embedded in a host matrix is analyzed theoretically. The important role of $d$ - electrons in formation of surface plasmon resonances is shown. A simple model is developed to describe the surface plasmons in nanorods. It is shown, that both for nanorods and spheroids the surface plasmon resonance wavelengths are linearly dependent on aspect ratios of these particles. The results of calculations are compared with experimental data for gold and silver particles and excellent agreement is demonstrated.

## 1. Introduction

The peculiar optical properties of metallic nanoparticles (MNP) are of great interest for a range of optical applications due to selective optical absorption in visible region conditioned by coherent collective oscillations of electrons - surface plasmons (SP) [1]. MNPs are already being used for surface enhanced Raman scattering (SERS) [1,2,3] and are promising candidates for application as nonlinear optical media in all-optical switching devices [4]. The simplest case allowing analytical description of SP spectrum is the model of ideal spherical MNP with size less than the wavelength, corresponding to the SP frequency. Indeed, under these conditions we have the limiting case of widely used Mie theory [5], which provides the well known expression for the SP frequency

$$\omega_{Mie} = \frac{\omega_p}{\sqrt{1+2\varepsilon_a}}, \qquad (1)$$

where $\omega_p^2 = 4\pi n_e e^2/m$ determines the plasma oscillations frequency in bulk, $n_e$ is the electron concentration, $m$ and $e$ are the electron mass and charge correspondingly, $\varepsilon_a$ is the real part of dielectric constant of matrix. Note that expression (1) describes a triple degenerate dipole oscillation. For such geometry the nonlinear optical phenomena [6], SP line broadening [7], radiation damping of SP oscillations and interaction with matrix electrons [8] are well investigated theoretically.

Very often the MNPs are being synthesized in the shape of nanorodes and spheroids, which show two SP bands corresponding to their axial symmetry [9-13]. The shape of a MNP influences essentially the processes of SP spectral line formation and oscillation decay and dephasing [14], interaction of electrons in MNPs with matrix electrons due to the concentration of the electric field around the tips of the particle [15] and radiation damping [2,12]. The influence of nonspherisity on optical properties of MNPs using phenomenological approach was investigated in [2,16] in order to identify the limiting factors of enhancement of SERS. The nonsphericity can strongly affect on broadening of SP lines caused by the interaction with matrix, which even for perfect spheres plays a significant role [8]. A numerical approach to the problem of finding the complex susceptibility of small particles and, consequently the absorption spectrum was developed in [17].

The purpose of this paper is to develop an approach for calculating the SP frequencies in noble metal nanosize spheroids and rods that will allow establishing a link between the shape parameters of MNPs and their SP frequencies.

## 2. The role of $d$ – electrons in formation of SP resonance

The expression (1) accounts for only free-electron contribution into the SP frequency of spherical particle embedded in dielectric media and does not account for contribution of $d$-electrons. However in noble metals, where the observed SP frequencies of MNPs are close to the $d$-electrons excitation threshold [7, 13, 18, 19], they play an important role in the dispersion of dielectric function. Thus, the analysis of the experimental data on absorption spectra of noble metal particles should be based on this peculiarity.

We use the model of electron gas with dielectric function $\varepsilon_s(\omega)$ embedded in medium with dielectric function $\varepsilon_d(\omega)$, which accounts for the optical response of $d$-bands. By additivity of susceptibilities the dielectric function of such composite system of $s$- and $d$-electrons is $\varepsilon_{sd} = \varepsilon_s + \varepsilon_d - 1$ (see for example [12]). Substituting the well-known function $\varepsilon_s(\omega) = 1 - \omega_p^2/\omega^2$, we obtain the following expression

$$\varepsilon_{sd}(\omega) = \varepsilon_d(\omega) - \omega_p^2/\omega^2, \qquad (2)$$

which will be used below in solving the electrostatic boundary problem. In fact $\varepsilon_{sd}$ is the dielectric function of a bulk noble metal that can be found by reflection and transmission measurements [20]. The following important characteristics of SP resonance can be determined from the independent experiments – the frequency, the value of $\varepsilon_{sd}$ at that frequency and the shape parameter. The theory developed below establishes a link between them and allows determining one of them with two other known.

The role of $d$-electrons is especially important for the frequencies at which $\varepsilon_d(\omega)$ is large. To prove this statement we consider Laplace`s equation for the electric field potential $\varphi$

$$\nabla^2 \varphi = 0, \qquad (3)$$

with the boundary conditions

$$\Phi = \Psi, \qquad \varepsilon_a \frac{\partial \Phi}{\partial n} = \varepsilon_{sd}(\omega) \frac{\partial \Psi}{\partial n}, \qquad (4)$$

where $\Psi$ and $\Phi$ are the inner and outer solutions of (3) respectively. In case of a noble metal sphere the boundary problem (3)-(4) leads to the equation $\varepsilon_{sd}(\omega) = -\frac{l+1}{l}\varepsilon_a$, $l = 1, 2, 3...$ and correspondingly the SP frequency of $l$-th oscillation mode can be presented as follows

$$\omega_l = \frac{\omega_p}{\sqrt{\varepsilon_d + \frac{l+1}{l}\varepsilon_a}}, \qquad (5)$$

As the noble metals in the range of $\sim 2 \div 4\,\text{eV}$ have large values of $\varepsilon_d$ [20] (much more than $\varepsilon_a$ of commonly used matrices), it follows from (5) that the frequencies of adjacent modes - $l$ and $l+1$ are close to each other, and even small nonsphericity causes the mixing of all multipolar frequencies. This is the reason of large differences between the SP frequencies of ellipsoidal and spherical noble metal particles (see e.g. [13] ). For the comparison we note that the SP frequencies of $2^l$-polarity mode of an ideal sphere when d-electrons are not involved in the transitions are [5]

$$\omega_l = \frac{\omega_p}{\sqrt{1 + \frac{l+1}{l}\varepsilon_a}}, \qquad (6)$$

and it is obvious, that the frequency separation is larger than in previous case.

In the next two sections we consider two types of nonspherical nanoparticles – rods and spheroids. While the spheroids allow the exact solution, there is no theory for nanorods. The approach developed below is applicable to the most important case of long nanorods, when as we show only the dipole oscillations are essential.

The influence of $d$-electrons is relatively less important for bulk metals as $\varepsilon_d(\omega_p)$ at plasma frequency is close to unity [20], and the observed value of $\omega_{Mie}$ only slightly differs from (1). However in the range of frequencies $2 \div 4$ eV the frequency dependence of $\varepsilon_d$ is essential and the Eq. (5) must be used instead of (1).

### 3. SP frequencies of nanorods

The influence of shape on the SP spectrum in Au nanorods was investigated experimentally in [9,18,19,21]. Two SP frequencies - $\omega_T$ and $\omega_L$ corresponding to transversal and longitudinal oscillations respectively were detected in all experiments. In order to calculate $\omega_T$ we adopt the approximation of infinite cylinder and solve Laplace`s equation (3) in cylindrical coordinates $\rho$ and $\vartheta$, assuming that $\varphi$ is independent on $z$ coordinate

$$\frac{\partial^2 \varphi}{\partial \rho^2} + \frac{1}{\rho}\frac{\partial \varphi}{\partial \rho} + \frac{1}{\rho^2}\frac{\partial^2 \varphi}{\partial \vartheta^2} = 0 \qquad (7)$$

The following boundary conditions at the surface of the cylinder must be satisfied

$$\Phi = \Psi, \qquad \varepsilon_a \frac{\partial \Phi}{\partial n} = \varepsilon_{sd}(\omega)\frac{\partial \Psi}{\partial n}. \qquad (8)$$

Here $\Psi$ and $\Phi$ are the inner and outer solutions of (7) respectively, which can be represented as follows

$$\Psi = \sum q_m \rho^m \left( A_m e^{im\vartheta} + B_m e^{-im\vartheta} \right), \qquad \Phi = \sum p_m \rho^{-m} \left( A_m e^{im\vartheta} + B_m e^{-im\vartheta} \right) \qquad (9)$$

From (8) and (9) we obtain

$$q_m R^m = p_m R^{-m}, \qquad -m\varepsilon_a p_m R^{-m-1} = m\varepsilon_{sd}(\omega) q_m R^{m-1}, \qquad (10)$$

where $R$ is the radius of the cylinder. Then the equation for determining the transversal frequency $\omega_T$ takes the form

$$\varepsilon_{sd}(\omega_T) + \varepsilon_a = 0, \qquad (11)$$

4which, using (2), can be also represented as

$$\omega_T = \frac{\omega_p}{\sqrt{\varepsilon_d(\omega_T) + \varepsilon_a}}. \qquad (12)$$

Thus, just in case of spherical MNP (see (1)), $\omega_T$ does not depend on angular quantum number and the cylinder radius. We note, that the approximation of infinite cylinder adopted above is supported by the experimental fact, that the transversal frequencies of nanorods are either not sensitive to the aspect ratio $\eta = L_0/2R$ ($L_0$ is the length of rod) [19], or demonstrate very weak dependence on it at $\eta \geq 2$ [18].

In [18] for the chosen values of $\varepsilon_a = 4$ and $\eta = 4.1$ the wavelength of the transversal plasmon oscillation at $\lambda_T = 520$ nm was observed. The absorption maximum at the same wavelength – 520 nm for the transversal oscillations of the surfactant coated nanorods in water has been observed for different aspect ratios - $\eta = 2.1 \div 5.5$ in [19] as well. Substituting the experimental values of $\varepsilon_{sd}$ from [20] into (11) we obtain exactly the same value for $\lambda_T$, which confirms the validity of our approach.

The longitudinal frequencies of nanorods can be estimated by the following reasoning. Consider very long metallic cylinder ($\eta \gg 1$), in which the electron cloud is slightly displaced along the axial direction. The potential energy of interaction of the "spilled-out" charges appearing at the edges of the cylinder is $W_{pot} = q^2 (\varepsilon_a L_0)^{-1}$, where $q$ is the charge at the edge. Since for $\eta \gg 1$ the electric field is mainly concentrated outside of the cylinder, only $\varepsilon_a$ enters into this expression. The "spilled-out" charge occupies the volume $V = S'x$, where $S'$ is the surface over which the charge is smeared, and $x$ is the thickness of the layer of smeared charge, $S'$ being larger than the cylinder cross section $S$ because of the repulsive forces. Thus the potential energy takes the form

$$W_{pot} = \frac{e^2 (kS)^2 n_e^2}{\varepsilon_a L_0} x^2, \qquad (13)$$

where $k = \frac{S'}{S}$ is the fitting parameter. On the other hand (13) can be represented as the potential energy of a harmonic oscillator with total mass $SL n_e m$ and frequency $\omega_L$, i.e. $W_{pot} = \frac{1}{2} S L_0 n_e m \omega_L^2 x^2$. Finally, comparing this expression with (13) we obtain the longitudinal wavelength as follows

$$\lambda_L = \frac{1}{k} \eta \sqrt{8\varepsilon_a} \lambda_p, \qquad (14)$$

where $\lambda_p = 2\pi c/\omega_p$, $\omega_p$ of bulk Au, according to [20] is 8.9 eV. Revealed linear dependence of longitudinal wavelength on the aspect ratio (14) has been detected in [9,18,21], however, as the experiments show, the straight line describing $\lambda_L$ dependence on $\eta$ has an intercept. Formula (14), in fact represents only the main term in the asymptotic expansion of function $\lambda_L(\eta)$

$$\lambda_L(\eta) \approx \frac{1}{k} \eta \sqrt{8\varepsilon_a} \lambda_p + a_0 + a_1 \eta^{-1} + a_2 \eta^{-2} + \ldots \qquad (15)$$

where $a_0$, $a_1, a_2$ ... are constants, $a_0$ being the intercept. In order to determine the value of our fitting parameter $k$ we equate the coefficient of $\eta$ in (14) to the slope $\mu \approx 1.137$ of the plot in the Fig. 6a of Ref. [18], which gives $k \approx 5.78$ for $\varepsilon_a = 4$.

In the experiment [19] for six ensembles of high quality samples with relatively small size dispersion of aspect ratios $\eta = 2.23, 2.81, 3.29, 3.34, 4.05, 4.75$ the dependence of $\lambda_L(\eta)$ is measured. The data obtained fit an increasing broken line with two jogs at $\eta = 3.29$ and $3.34$. Note that for these two values the dependence of the average period of breathing vibrational mode on $\eta$ also displays two jogs. To obtain an average slope $\mu$ for $\lambda_L(\eta)$ in this experiment we apply the least-square method and find for $\mu \approx 1.121$.

Since the electric field of long enough nanorod is mainly concentrated in the matrix material, it is clear from the physical point of view that the fitting parameter $k$ depends on effective dielectric constant of matrix and coating. For this reason, it is appropriate to introduce a new fitting parameter of the theory - $K = \sqrt{8\varepsilon_a}/k$ which for the given $\lambda_p$ determines the slops of [18] and [19]. In fact the small difference in slopes in these two experiments is conditioned by not equal values of $K$, which is a consequence of differences in the surfactant coatings.

### 4. SP frequencies of spheroidal MNPs

In contrary to the case of nanorods the theoretical background of SP spectrum of spheroidal particles is well worked out. The point is that the electrostatical boundary problem for ellipsoid allows a simple analytical solution [22], the limiting case of which is a solution for spherical MNPs.

Often (see [18,21]) the interpretation of the results of the experiments on nanorods is being done by the involvement of formulas for spheroids [22, 23]. Using the known solution of boundary problem for ellipsoid (Eq. (8,12) of Ref. [22]) in case of absence of external field, we obtain

$$\varepsilon_{sd}(\omega) + \frac{1 - n_{xy,z}}{n_{xy,z}} \varepsilon_a = 0 \quad (16)$$

where $n_{xy,z}$ is the depolarization factor, which is determined through the aspect ratio $\eta$ and eccentricity $e$ as it follows from [50]

$$n_z = \frac{1-e^2}{e^3}(\mathrm{Arth}(e) - e), \quad n_x = n_y = \frac{1}{2}(1 - n_z), \quad e = (1 - \eta^{-2})^{1/2} \quad (17)$$

From these formulas we obtain in case of strongly prolate spheroid ($\eta \gg 1$, $n_x = n_y$) the following expression for the longitudinal SP wavelength

$$\lambda_L(\eta) \approx \frac{\eta}{\sqrt{\ln \eta}} \sqrt{\varepsilon_a} \lambda_p. \quad (18)$$

This dependence of $\lambda_L$ on $\eta$ for $2 \leq \eta \leq 5$, as it can be shown by plotting (18) only slightly differs from the linear dependence. Thus Eq.(18) and the first two terms in the right hand side of (15) express the same behavior. The fit of (18) to linear part of (15) gives

$$\lambda(\eta) = \left(\frac{\eta}{1.95} + 1.35\right)\sqrt{\varepsilon_a}\lambda_p \quad (19)$$

It is easy to see that the slopes of (14) and (19) are very close to each other. Thus, our approach based on clear physical arguments leading for the nanorods to (14) and (15) describes with high accuracy the spectrum of prolate spheroids as well.

In order to clarify the role of $d-$ electrons in the formation of SP resonance in nanorods we use formulas (2) and (16) to present $\omega_L$ for the ellipsoid with arbitrary aspect ratio

$$\omega_L = \frac{\omega_p}{\sqrt{\varepsilon_d - \left(1 - \frac{1}{n_z}\right)\varepsilon_a}} \qquad (20)$$

This formula demonstrates an important peculiarity of noble metal MNPs. Namely, the longitudinal frequency in case of strongly prolate ellipsoids ($n_z \ll 1$) is determined by the host matrix dielectric constant $\varepsilon_a$, whereas for strongly oblate ones ($n_z \approx 1$) the main contribution is conditioned by $\varepsilon_d$. Thus, along with increase of aspect ratio $\eta$ the dominant role in formation of longitudinal SP resonance smoothly passes from $\varepsilon_d$ to $\varepsilon_a$.

Experimentally the silver spheroidal particles were investigated by time-resolved optical pump-probe spectroscopy [13]. In this work the extinction spectra of spheroidal silver nanoparticles with mean axis length $a = b \approx 40\,\text{nm},$ and $c \approx 100\,\text{nm}$ for two light polarizations were measured and the long-axis and short-axis SP resonances were detected at $\hbar\omega_{la} = 1.88\,\text{eV}$ and $\hbar\omega_{sa} = 3.06\,\text{eV}$. In the same paper the SP resonance at 2.82 eV in the spherical particle embedded in dielectric was reported as well.

To apply to this experiment the theory developed for the spheroids we note, that the observed frequency of 2.82 eV is a root of Eq.(16) for the sphere, e.g.

$$n_z = \frac{1}{3}, \quad \text{and} \quad \omega_L = \omega_T = \omega_{sp}, \qquad (21)$$

which brings to the following relationship between $\varepsilon_{sd}(\omega_{sp})$ and $\varepsilon_a$

$$\varepsilon_{sd}(\omega_{sp}) + 2\varepsilon_a = 0. \qquad (22)$$

Taking the value of $\varepsilon_{sd}(\omega_{sp}) = -6.549$ from [20], we determine from (22) the dielectric constant of the host matrix as $\varepsilon_a \approx 3.275$.

In order to compare the results of the measurements with the theory it is necessary to check in which extent the formulas for spheroids (16) and (17) correspond to the aspect ratio $\eta \approx 2.5$ and observed frequencies 1.88 eV and 3.06 eV. Taking the values of $\varepsilon_{sd}(\omega)$ at these frequencies from [59] we find from (16), that $n_{xy} = 0.41$, and $n_z = 0.14$. Substituting these values into (17) we obtain for the aspect ratio correspondingly $\eta \approx 2.43$, which differs from the measured value only by 3%. Thus, the SP oscillation spectrum of silver spheroidal MNPs, can be calculated by the formulas (16) and (17).

It is interesting to solve also the inverse problem which is the following: using the Eqs. (16) and (17) to find the long-axis and short-axis SP frequencies starting from the given aspect ratio value $\eta = 2.5$ and $\varepsilon_a = 3.275$. The calculations give: $n_z = 0.135$, $\varepsilon_{sd} \approx 20.984$, , and $n_{xy} = 0.433$, which lead to the following values of required frequencies: $\hbar\omega_{la} \approx 1.848\,\text{eV}$ instead of observed 1.88 eV, and $\hbar\omega_{sa} \approx 3.12\,\text{eV}$ instead of 3.06 eV.

Since the observed SP energy of spherical silver particle in vacuum ($\varepsilon_a = 1$) is 3.5 eV [12], at this frequency we obtain from (22) $\varepsilon_{sd} = -2$. It is interesting to note, that the experiment [20] for bulk silver at 3.5 eV gives exactly the same value for $\varepsilon_{sd}$.

## 5. Conclusion

Experimentally revealed linear dependence of SP resonance wavelength on aspect ratio of nanorods and spheroids is justified theoretically. The coincidence of the observed and calculated SP frequencies of noble metal nanorods and spheroids shows, that it is possible to describe with high degree of accuracy the optical properties of MNPs developing a simple physical models. This allows extracting the shape parameters of MNPs from the results of absorption measurements.

The excellent agreement between experimental data on SP spectrum of MNPs [13,18,19] and results of calculations using bulk optical constants [20] indicates that noble metal particles of size down to 10 nm are still described by the bulk dielectric function.